\documentclass[12pt]{iopart}
\usepackage{graphicx}

\usepackage{xspace}
\newcommand{\ket}[1]{\ensuremath{|#1\rangle}\xspace}
\newcommand{\bra}[1]{\ensuremath{\langle #1|}\xspace}
\newcommand{\op}[2]{\ensuremath{| #1 \rangle \langle #2 |}}

\let\vect=\overrightarrow

\begin{document}

\title{Optimizing H1 cavities for the generation of entangled photon pairs}
\author{M. Larque, T. Karle, I. Robert-Philip and A. Beveratos}
\address{Laboratoire de Photonique et Nanostructures LPN-CNRS UPR-20, Route de Nozay, 91460 Marcoussis, France}
\ead{Alexios.Beveratos@lpn.cnrs.fr}

\begin{abstract}
We report on the theoretical investigation of photonic crystal
cavities etched on a suspended membrane for the generation of
polarization entangled photon pairs using the biexciton cascade in
a single quantum dot. The implementation of spontaneous emission
enhancement effect increases the entanglement visibility, while
the concomitant preferential funneling of the emission in the
cavity mode increases the collection of both entangled photons. We demonstrate and quantify
that standard cavity designs present a polarization dependent emission diagram, detrimental to entanglement. The optimization of H1 cavities allows to obtain both high collection efficiencies and polarization independent emission, while keeping high Purcell factors necessary for high quality entangled photon sources.
\end{abstract}

 \submitto{\NJP} \maketitle

Entangled photon sources play an important role in quantum
communication networks or quantum information processing
\cite{Shor97, Grover97}. For the former, they are a necessary
resource for the realization of quantum repeaters \cite{Collins05}
based on quantum teleportation or quantum entanglement swapping.
In the first demonstrations of such relays, parametric down
conversion sources have been used for the generation of entangled
photon pairs \cite{Ursin04, Landry07,Pan98, Halder07}. Such
non-linear sources of entanglement can combine narrow spectral
bandwidths with a maximal generation rate \cite{Halder07, Kumar05,
Fulconis07}. Although these sources may be very useful and easy to
implement, they always suffer from the Poissonian statistics of
the emitted photons pairs leading to multipair emission and thus
decreasing the fidelity of entanglement \cite{Scarani05}. Being
able to produce polarization entangled photon pairs on demand
would be an important step towards robust quantum relays. Such
sources can be obtained from the biexciton-exciton cascade
emission of a single quantum dot \cite{Benson2000}, and first
experimental demonstrations have been reported \cite{Akopian2006,
Stevenson2006}. Obtaining entangled photons pairs, however, from
such quantum dots sources with both high fidelity and high
collection efficiency remains a problem. Implementing Cavity
Quantum ElectroDynamics effects by embedding a single quantum dot
in a microcavity could not only improve the fidelity of the
emitted pair \cite{Larque2008} by taking advantage of the Purcell
effect, but also by enhance the collection efficiency
\cite{Barnes2002, Balet2007}. One promising microcavity for such
purpose is the single defect hole cavity in a triangular
lattice of holes (H1) etched on a suspended membrane, due
to its small mode volume and its polarization degeneracy. However,
in a standard H1 cavity, the radiation pattern of the two
fundamental degenerate modes do not overlap, leading to photon
distinguishability and thus destroying entanglement. Theoretical
calculations demonstrate that this radiation pattern can be
strongly modified by changing, for instance, the position of the
holes surrounding the defect \cite{Roemer2008}. This modification
of the design is necessary to avoid distinct emission patterns.
This emission pattern distinguishability is related to the mode
overlap.

In this paper, we report on the theoretical investigation of H1
photonic crystal cavities etched on slab membrane, in order to
obtain both high collection efficiencies for both photons and a
high overlap between the two fundamental energy-degenerate modes.
The dependency of the Bell inequalities as a function of the mode
overlap is derived. We also investigate the impact of the position
of the quantum dot inside the cavity on entanglement visibility
and collection efficiency.

\section{Entangled state density matrix for non overlapping modes}

Polarization entangled photon pairs can only be obtained if and
only if, even in principle, the polarization of the photon can not
be determined by measuring another degree of freedom as for
example the photon's energy. In the same way, if the emission mode
of one of the photons of the pair does not perfectly match the
emission mode of the other photon, the non-maximal overlap between
the two emission modes will reduce the fidelity of entanglement.
Our analytical derivation of this non-maximal mode overlap effect
is based on the density matrix of the photon pair emitted by the
cascade emission from the biexcitonic level of a single quantum
dot.

\begin{figure}[h]
   \begin{center}
   \includegraphics[height=5cm]{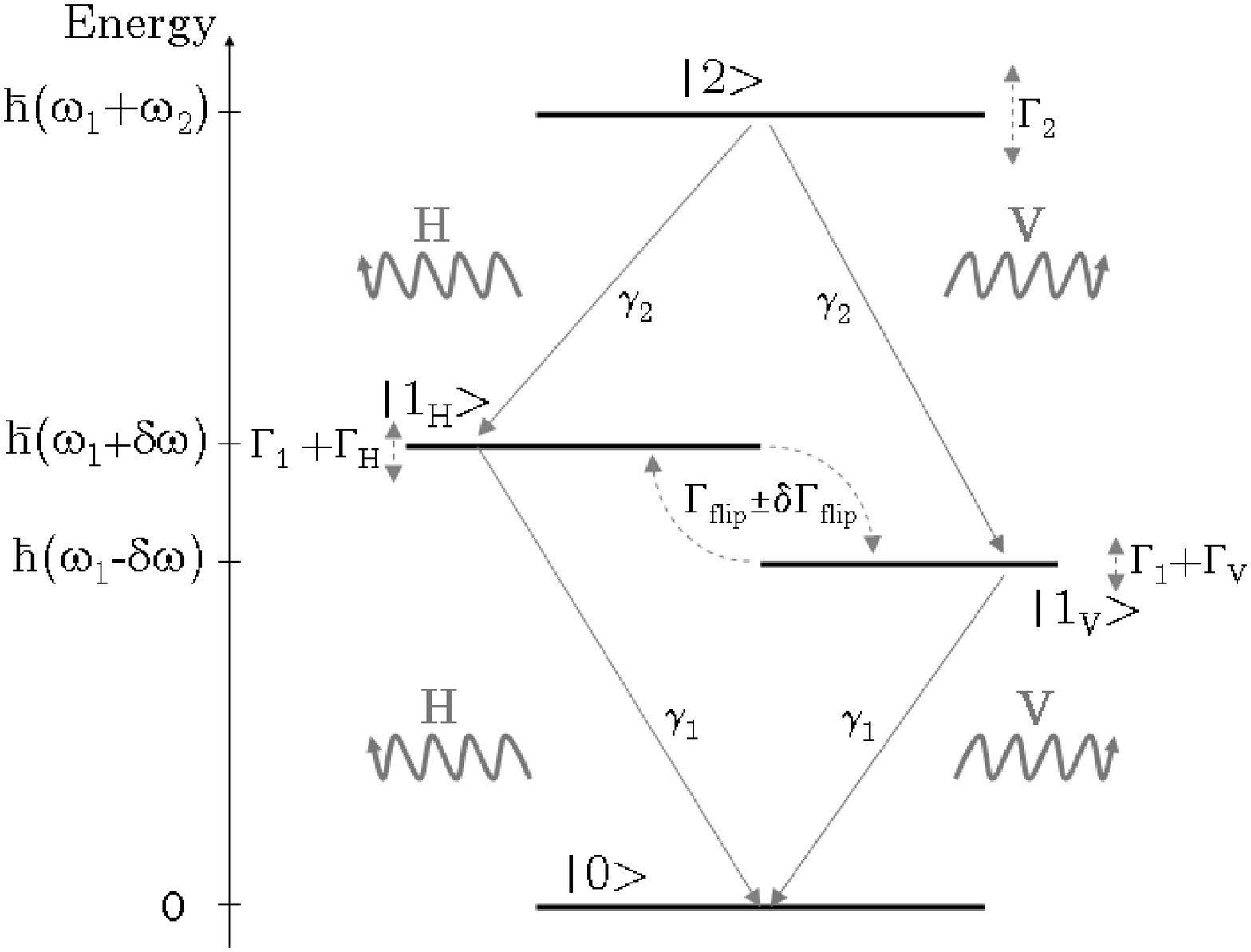}
   \caption{Schematic description of the two-photon cascade in a
typical quantum dot four-level system with an energy splitting
2$\hbar \delta\omega$ of the relay level, yielding two collinearly
polarized photons (either $H$ or $V$).}
   \end{center}
   \label{fig:cascade}
\end{figure}

The eigenbasis of the dot involves four levels : $\ket{2}$
(biexcitonic level), $\ket{1_H}$ and $\ket{1_V}$ (two excitonic
levels with opposite angular momenta) and $\ket{0}$ (fundamental
level). In this eigenbasis, the emitted photons are linearly
polarised along the horizontal (H) or vertical (V) directions. The
density matrix of the photon pair in this particular basis
$\mathcal{B}=[H_1H_2,H_1V_2,V_1H_2,V_1V_2]$ where the subscript
$i=1,2$ is related to the photon emitted by the biexcitonic level
and excitonic level respectively, can be written in the form
\cite{Larque2008}:
\begin{equation}
\overline\rho = \left( \begin{array}{cccc}
    \alpha & 0 & 0 & d - i c_1 \\
    0 & \frac{1}{2}-\alpha & c_2 & 0 \\
    0 & c_2 & \frac{1}{2}-\alpha & 0 \\
    d + i c_1 & 0 & 0 & \alpha
    \end{array}\right)
\end{equation}

\noindent where:
\begin{eqnarray}
\alpha &=& \frac{1}{2} \frac{\gamma_1+\Gamma_{flip}}{\gamma_1+2\Gamma_{flip}} \nonumber\\
d &=& \frac{1}{2} \frac{\gamma_1(\gamma_1+2\Gamma+\Gamma_{flip})}{(2\delta\omega)^2 + (\gamma_1 + \Gamma_{flip} + \Gamma)^2 - (\delta\Gamma_{flip})^2}\\
c_1 &=& \frac{1}{2} \frac{\gamma_1 \delta\omega}{(2\delta\omega)^2 + (\gamma_1 + \Gamma_{flip} + \Gamma)^2 - (\delta\Gamma_{flip})^2}\nonumber\\
c_2 &=& \frac{1}{2} \frac{\gamma_1
\delta\Gamma_{flip}}{(2\delta\omega)^2 + (\gamma_1 + \Gamma_{flip}
+ \Gamma)^2 - (\delta\Gamma_{flip})^2}\nonumber
\end{eqnarray}

\noindent with $\gamma_1$ the exciton decay rate. $\Gamma =
\Gamma_H+\Gamma_V$ is the cross-dephasing rate between the two
excitonic states. $\Gamma_{flip}\pm \delta\Gamma_{flip}$ describe
phenomenologically relaxation mechanisms between the two
excitation states, leading to incoherent population transfers
between these two states ($\delta\Gamma_{flip}$ takes into account
the possible rate asymmetry in this process). $\delta\omega$ is
the energy splitting of the excitonic levels. $\Gamma_1$ is the
pure dephasing rate induced by dephasing processes that occur
simultaneously and attach the same information to the phase and
energy of these two excitonic levels. $\Gamma_H$ and $\Gamma_V$
are polarization-dependent pure dephasing rates induced by
dephasing processes that do not  identically affect the two relay
levels and whose impact depends on the polarization of the
excitonic states.

We consider that the exciton and biexciton photons are emitted in
the same cavity mode, ie the cavity mode is resonant with both
transitions. The cavity mode is doubly-degenerate in
polarisation, due to the $C_6$ symmetry of the H1 cavity. We can
consequently describe the photons polarization in the ($H$, $V$)
basis defined previously, independently of the orientation of the
dot with regards to the orientation of the photonic crystal. We
define $\Phi_H(\vec{r})$ (resp. $\Phi_V(\vec{r})$) the complex
spatial far field distribution of the horizontal ($H$) (resp.
vertical ($V$)) polarization modes. Propagation occurs along the
orthogonal direction  to the photonic crystal membrane and
$\vec{r}$ denotes the radial vector perpendicular to the
propagation axis. The density matrix can be rewritten under the
form:
\begin{equation}
\rho(\vec{r_1},\vec{r_2})_{xy, uv} =
\Phi_x(\vec{r_1})^\dagger*\Phi_y(\vec{r_2})^\dagger *
\overline\rho_{xy, uv} * \Phi_u(\vec{r_1})*\Phi_v(\vec{r_2})
\label{eq:normNewRho}
\end{equation}

\noindent with $xy$ and $uv$ $\in$ $\mathcal{B}$ and $
\overline\rho_{xy, uv}$ being the density matrix element on line
$xy$ and column $uv$. $\rho(\vec{r_1},\vec{r_2})_{xy, uv}$ is the
density matrix element on line $xy$ and column $uv$ of the new
density matrix $\rho(\vec{r_1},\vec{r_2})$. Let $t(r)$ be the
function describing the detectors' active areas which are placed
along the propagation axis. There are in fact two distinctive
detectors (one for each photon of the pair \cite{Larque2008}) but
we suppose that they have the same sensitive area for the sake of
simplicity. The density matrix can be reduced for the detected
photon pairs to:
\begin{equation}
\rho = \frac{\int d^2r_1\ d^2r_2\ t(\vec{r_1}) t(\vec{r_2})
\rho(\vec{r_1},\vec{r_2})}{Tr(\int d^2r_1\ d^2r_2\ t(\vec{r_1})
t(\vec{r_2}) \rho(\vec{r_1},\vec{r_2}))} \label{eq:normalisation}
\end{equation}

We consider that the cavity is positioned at the focal point of a
microscope objective, which transforms the emitted far field into
the complex transverse shape of a propagative beam. In the first
order approximation, $\Phi_H$ and $\Phi_V$ are real and positive,
corresponding to the case where the transverse phase is constant
in the propagative modes (plane wave approximation). Let $k$ and
$e$ be:
\begin{eqnarray}
k = \int d^2r\ t(\vec{r}) \sqrt{\Phi_H(\vec{r}) \Phi_V(\vec{r})} \\
e = \int d^2r\ t(\vec{r}) \Phi_H^2(\vec{r}) = \int d^2r\
t(\vec{r}) \Phi_V^2(\vec{r})
\end{eqnarray}

\noindent The overlap factor $K$ can be expressed as $K=k^2/e^2$
and the final expression of the detected photon density matrix in
the case of non-maximal overlap of the two emission modes is:
\begin{equation}
\rho = \left( \begin{array}{cccc}
    \alpha & 0 & 0 & (d - i c_1) K \\
    0 & \frac{1}{2}-\alpha & c_2 K & 0 \\
    0 & c_2 K & \frac{1}{2}-\alpha & 0 \\
    (d + i c_1) K & 0 & 0 & \alpha
    \end{array}\right)
\end{equation}

\noindent Note that only the coherence terms are modified, and are
multiplied by the overlap factor $K$. When both modes do not
overlap ($K=0$), the mutual coherence is erased and entanglement
vanishes. On the contrary, maximally entangled states can only be
obtained for $K=1$. Following \cite{Larque2008}, Bell inequalities
can be rewritten as $S=2\sqrt{2}\left(\alpha+K*(d-c_2)\right) >
2$. Even in the case of a single dot emitting maximally entangled
photons, a minimum overlap of $K>2/\sqrt{2}-1=41\%$ is required in
order to violate Bell inequalities.

\section{H1 cavity for maximally entangled photons}

We consider the H1 cavity as potential candidate for the
generation of entangled photon pairs since it sustains two energy
degenerate dipole modes with a field maximum in the center of the
cavity. This cavity offers both a low mode volume and
theoretically high quality factors by fine tuning the inner holes
\cite{Roemer2008}.

\begin{figure}[h]
   \begin{center}
   \begin{tabular}{cc}
   \includegraphics[height=4cm]{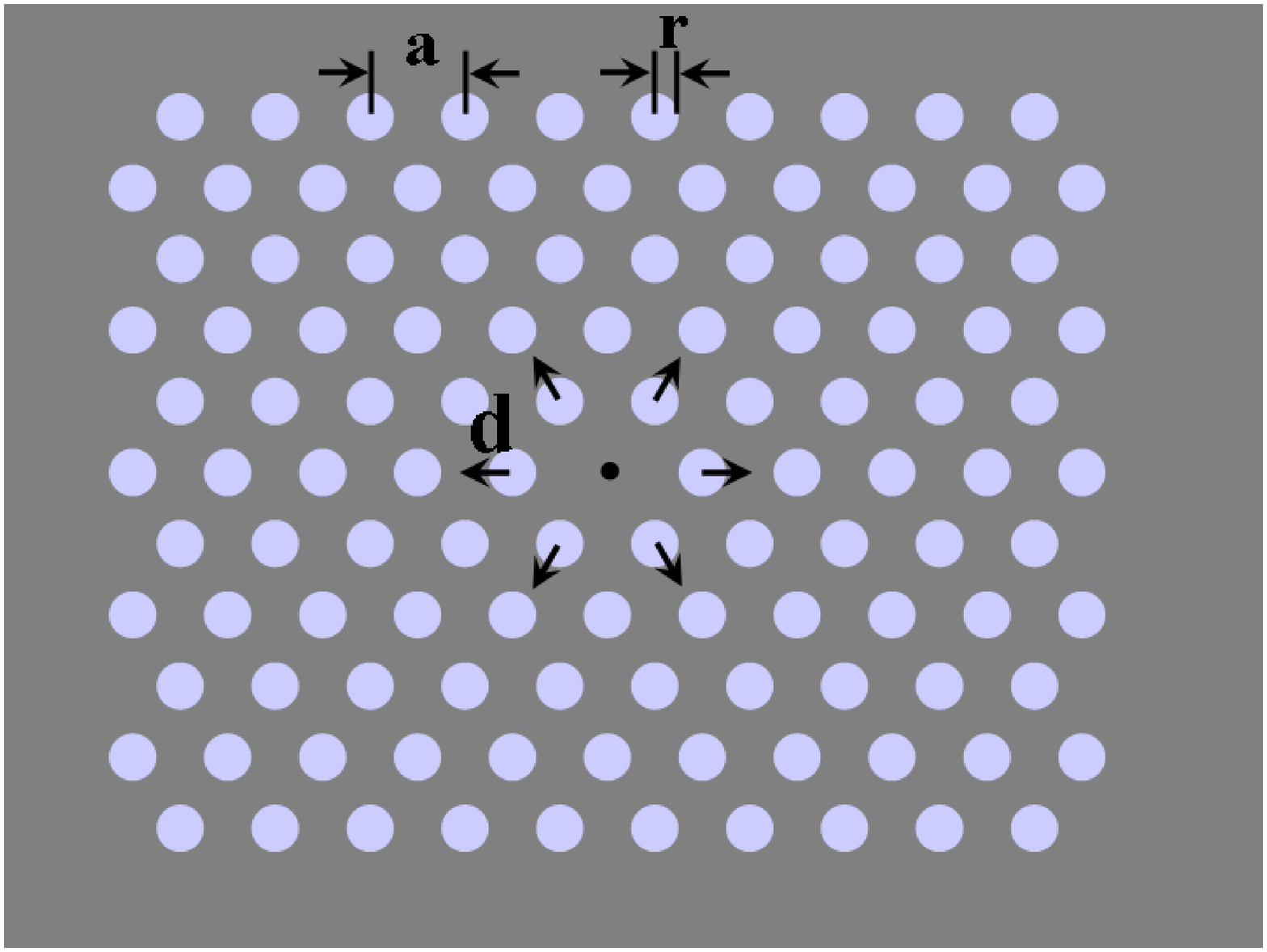} &
   \includegraphics[height=4cm]{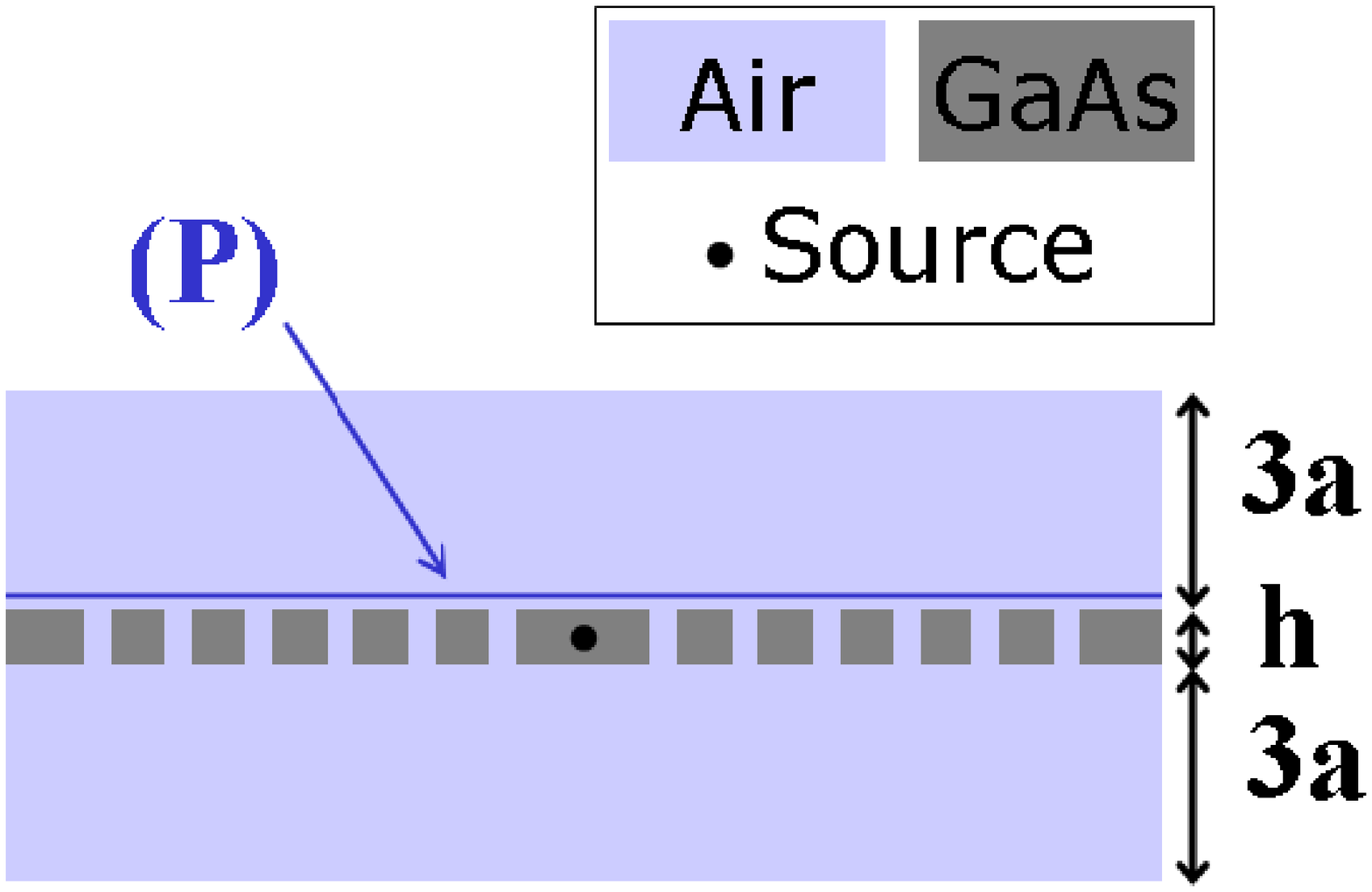} \\
   (a) & (b)
   \end{tabular}
   \end{center}
   \caption{H1 photonic crystal cavity. The inner holes are shifted by the quantity d from 0 (standard H1 design) to 0.18a. The blue line is the plane where the field is registered at the end of the simulation for radiative pattern calculations.}
   \label{fig:cavityH1}
\end{figure}

Simulations of H1 cavities were performed with the 3D
finite-difference time-domain (FDTD) method, using a freely
available software package with subpixel smoothing for increased
accuracy\cite{Farjapour2006}. The simulated structure is depicted
in figure \ref{fig:cavityH1}. The H1 photonic crystal (figure
\ref{fig:cavityH1}(a)) has a lattice constant equal to $a=$270 nm
and the holes have a radius of $r_h=$80 nm. The refractive index of the GaAs
membrane is equal to $n=$3.46. Above and below the membrane, a
free space volume is inserted, with a thickness of 3$a$. The
simulation volume is finally surrounded by Split Field Perfect
Matched Layers (PML). A temporally short Gaussian dipole pulse
(with a width of 10 optical oscillations) is launched in the
center of the cavity (figure \ref{fig:cavityH1}(b)) and used as a
white light source. After extinction of the source, the
electromagnetic field evolves freely over a time corresponding
to approximately 300 optical cycles of the source, after which
all low quality factor modes have radiated, thus leaving only the
desired cavity mode in the simulation volume. In such conditions,
the decay of the field amplitude at some fixed non-nodal point
inside the cavity follows a simple exponential function of rate
$\Gamma_c$. The emission wavelength is determined by measuring the
optical oscillation frequency.

The collection efficiency is defined as the ratio between the
incident power within a given emission cone normal to the
membrane, over the total emitted power. At any given time, the
total emitted power is given by $P_{ref} = \Gamma' * W$ where W
denotes the energy inside the cavity, and $\Gamma'$ is the
intensity decay rate ($\Gamma'=2\Gamma_c$). Let U be the total
energy in the simulated volume at time $t$. The energy outside the
cavity ($U_{out}$) corresponds to the energy emitted by the cavity
which has not yet reached the edge of the simulation volume,
$U_{out}=\Gamma'*W*D/c$ with $D$ being the radius of the simulated
volume and $c$ the speed of light. Thus $U=W*(1+D/c)$. Since $D$
is only a few micro-meters wide, $D/c << 1$, and the total emitted
power can be written as $P_{ref}=2\Gamma_c U$.

\begin{figure}[h]
   \begin{center}
   \includegraphics[height=6cm]{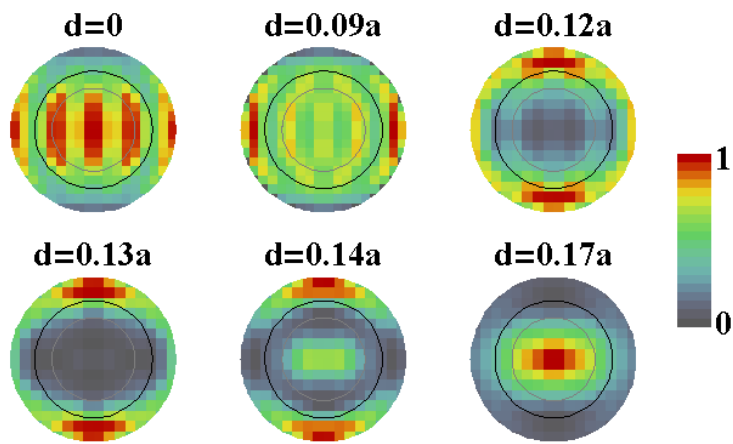}
   \caption{Emission patterns for various hole displacement $d$ for a membrane thickness of $h=$0.26$\mu$m. Each pattern is normalized to its maximum value. The distance to the center is $sin(\theta)$ where $\theta$ is the normal angle to the membrane. The gray (resp. black) circle represents an objective of numerical aperture NA=0.5 (resp. NA=$sin(\pi/4)=0.7$)}
   \label{fig:Emission}
   \end{center}
\end{figure}

The emission mode of the cavity is estimated following reference
\cite{Vuckovic2002} (mainly Eq. 23). This method relies on the
complex value of the electromagnetic field on a plane ($P$) just
above the membrane (see figure \ref{fig:cavityH1}(b)) at some time
(in our case the end of the simulation). The real part of the
electromagnetic field is directly measured on ($P$), and the
imaginary part is deduced from measurements of the real part of
the electromagnetic field a quarter oscillation later (at the
cavity's resonant frequency), taking into account the losses
induced during this quarter of cycle. This allows us to extract
the far field emission mode from the light-cone of the spatial
Fourier transform of the field with a unique simulation run in
real values, thus saving valuable calculation time. Emission
patterns for a membrane thickness of $h=0.26 \mu m$ as a function
of the hole displacement $d$ are depicted in figure
\ref{fig:Emission} and correspond qualitatively with the emission
patterns calculated by Roemer \textit{et al} \cite{Roemer2008}
using a 3D Finite Element Maxwell Solver. For small hole
displacements ($d \leq 0.10$), the emission diagram is almost
spherical, whereas by increasing the hole displacement ($d \geq
0.14$), a pronounced, directional Gaussian-like, central peak
appears.

\begin{figure}[h]
   \begin{center}
   \includegraphics[height=3.9cm]{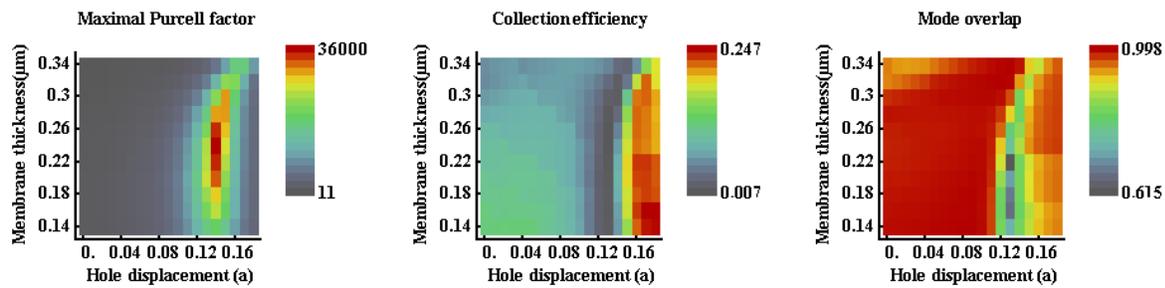}
   \caption{Maximal Purcell factor, collection efficiency, and mode overlap as a function of the hole displacement ($d$, in cristal units $a$) and membrane thickness ($h$, in $\mu m$).}
   \label{fig:scan2D}
   \end{center}
\end{figure}

\begin{figure}[h]
   \begin{center}
   \includegraphics[height=3.4cm]{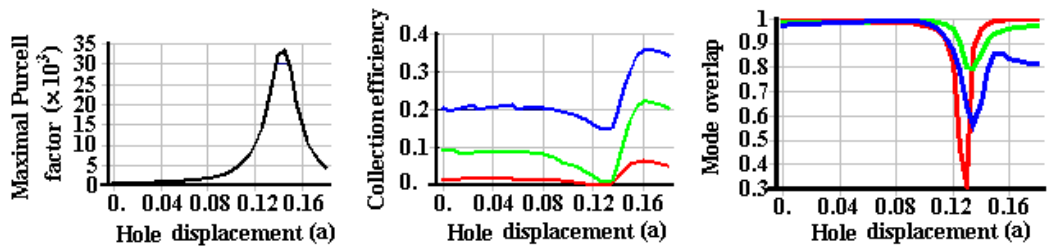}
   \caption{Maximal Purcell factor, collection efficiency, and mode overlap as a function of the hole displacement ($d$, in cristal units $a$) for a membrane thickness $h=0.26\mu m$. The collection efficiency and the mode overlap are calculated for numerical apertures NA=0.2 (red curves), NA=0.5 (green curves) and NA=0.7 (blue curves).}
   \label{fig:scan1D}
   \end{center}
\end{figure}

A systematic analysis of the maximal Purcell factor ($F_p^{max}$),
collection efficiency ($\eta$), and mode overlap ($K$) has been
performed by varying two parameters: the hole displacement $d$ and
the membrane thickness $h$. Figure \ref{fig:scan2D} summarizes the
results where the mode overlap and collection efficiency are
estimated for an objective with a numerical aperture of 0.5. By
maximal Purcell factor, we mean the Purcell factor obtained for a
resonant punctual monochromatic source placed at the maximum of
the electrical field intensity. A more detailed view of the
variation of these parameters as a function of the hole
displacement is given in figure \ref{fig:scan1D}, in which the
membrane thickness is fixed to $h=0.26\mu m$ and three numerical
apertures have been taken into account (0.2, 0.5 and 0.7). The
wavelength $\lambda_c$ of the cavity depends strongly on both
parameters $d$ and $h$. A good linear approximation is $\lambda_c
= d \times 0.28\mu m + h \times 0.69 + 0.82\mu m$ with 4 nm of
maximal deviation. A global homothetic transformation of the
design, including the membrane thickness, should be latter applied
to tune the cavity to the desired wavelength but is not taken into
account here.

As we demonstrated earlier \cite{Larque2008} a Purcell factor of
$10$ should be sufficient to restore entanglement in the emission
cascade from single quantum dot. From this point of view alone,
the whole domain of variation of the two parameters studied here
satisfies this condition. However, spontaneous emission
enhancement of the exciton rate is not sufficient to realize a
deterministic efficient entangled photon source: efficient
coupling of both photons to the cavity mode, high emission mode
overlap and high collection efficiency are also needed. Let us
first consider the coupling of both photons to the cavity mode.
Exciton and biexciton lines are usually separated by about $2nm$.
If the exciton is resonant with the cavity in order to obtain the
highest spontaneous emission acceleration, the biexciton should
preferably be also quasi-resonant with the cavity mode, in order to
benefit from the redirection of the emission and to increase the
collection efficiency of the biexcitonic photon. If we set a
minimum Purcell factor for the quasi-resonant biexcitonic line to
a value of about $5$, this in turn limits the Purcell factor which
can be reached for the excitonic line to a maximum value of about
$110$ for a cavity of modal volume $0.7 (\lambda_c/n)^3$. This
limitation excludes a large domain of parameters value around the
hole shift of $d=0.13$. Let us consider now the problem of
emission mode overlap and collection efficiency.  Since the
highest collection efficiencies with a Gaussian-like emission
pattern are reached for a membrane thickness of $h=0.26\mu m$, all
further discussions are with this fixed value.

At low hole displacement ($d\leq0.09a$), high mode overlap can be
obtained (in excess of 95\%) but at the expense of a low
collection efficiency (below 10\% for standard microscope
objectives (NA$\leq0.5$) ) and does not exceed 22\% for high
numerical aperture objectives (NA=0.7) with a theoretical maximum
of 50\% since light is emitted upwards and downwards with the same
intensity. The emission diagram (figure \ref{fig:Emission}) for a
hole displacement of d=0.1a clearly explains the situation. The
mode is almost uniform in every direction giving rise to a high
mode overlap, and the collection efficiency scales as the
objective's numerical aperture. In the range $0.11a\leq d\leq
0.14a$ the mode overlap presents a distinct dip down to 57\% for
an objective with NA=0.7, and the collection efficiency drops to
the same extent. In this region, the Purcell effect reaches its
maximum (d=0.145a). The drop in the collection efficiency is
correlated to the increase of the quality factor, corresponding to
a better confinement of the light inside the photonic crystal slab
and a reduction of vertical losses at the $\Gamma$ point. For
larger hole displacements ($d\geq0.15a$) the collection efficiency
increases sharply reaching 22$\%$ for a NA=0.5 (and $d=0.16a$),
corresponding to a 4-fold increase compared to a hole
displacement of $d=0.10a$. At the same time, the mode overlap
increases up to 96\% reaching almost the values obtained at low
values of d. The mode profile (fig \ref{fig:Emission}) is almost
TE$_{00}$ in the propagation direction perpendicular to the
membrane. A numerical aperture of NA=0.2 increases the overlap up
to almost 100\% but at the expense of a low collection efficiency.
On the other hand, a numerical aperture NA=0.7 increases the
collection efficiency by a factor of 1.5 compared to a numerical
aperture NA=0.5, but the mode overlap does not exceed 83\%
indicating that almost half of the energy is astigmatic. As a
conclusion, a hole displacement of $d=0.16a$ with an objective
with numerical aperture of NA=0.5 and membrane thickness
$h=0.82\lambda_c/n$ seems to be the optimum in terms of collection
efficiency (of the order of 22$\%$ with a maximal calculated value of 50$\%$) 
and mode overlap (of the order of 96$\%$).

\section{Impact of the position of the dot in the cavity}

Until now the quantum dot has been considered to be perfectly placed in the center of the cavity, implying that both polarizations undergo the same Purcell effect and that the cavity mode is equally fed for both polarizations. Deterministically aligning a photonic crystal around a single quantum dot so that the dot is positioned in the center of the cavity is technologically extremely challenging but mandatory. Several techniques are being developed \cite{Badolato2005, Gogneau2008} although due to experimental uncertainties which are essentially due to the electronic beam lithography process, the mismatch of the quantum dot position with respect to the center of the H1 cavity can be up to 10 nm.  

\begin{figure}[h]
   \begin{center}
   \begin{tabular}{ccc}
   \includegraphics[height=2.5cm]{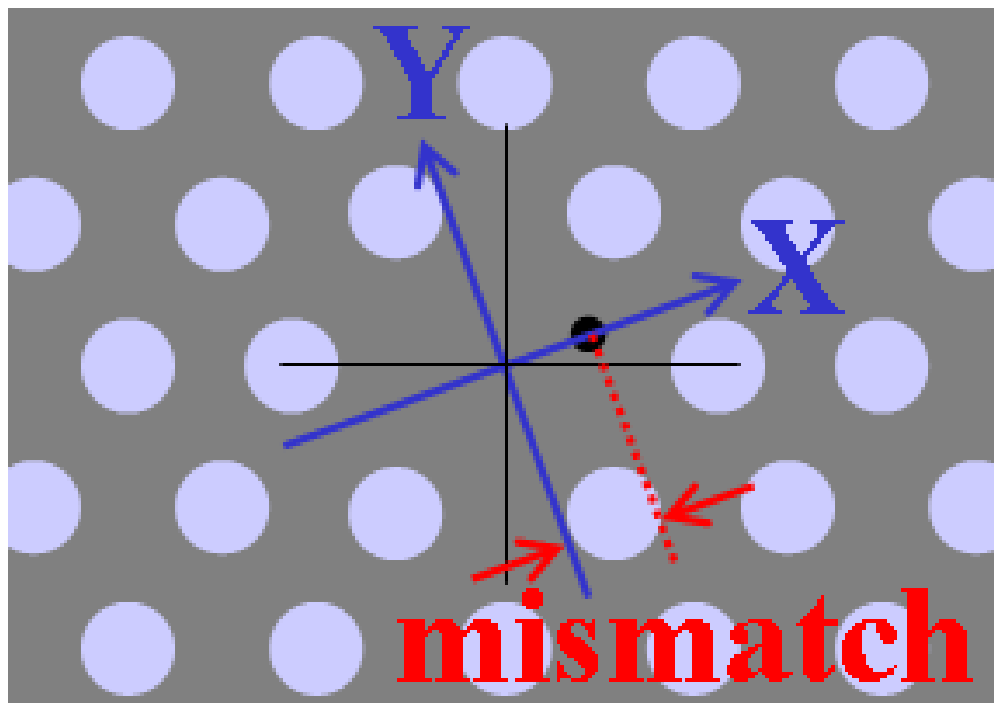} &
   \includegraphics[height=3.2cm]{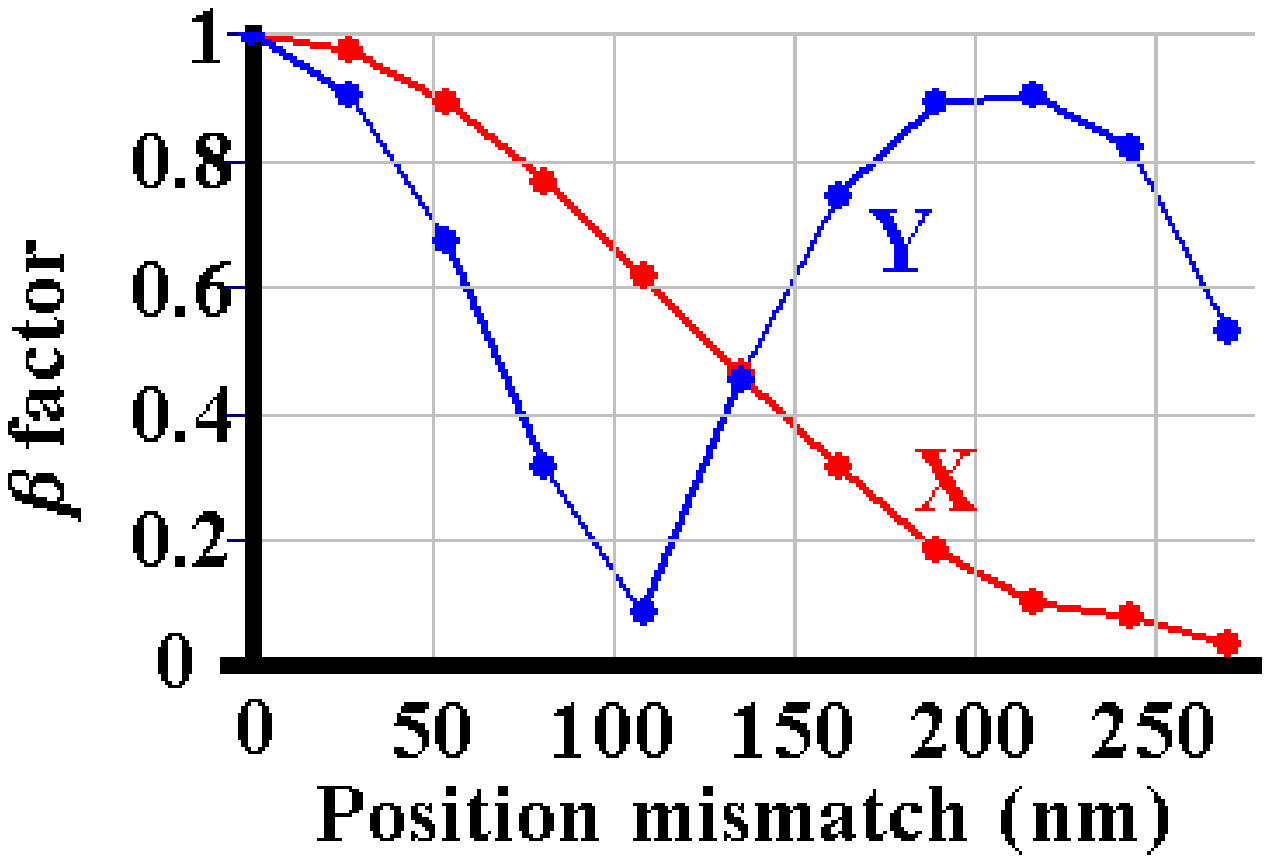} &
   \includegraphics[height=3.2cm]{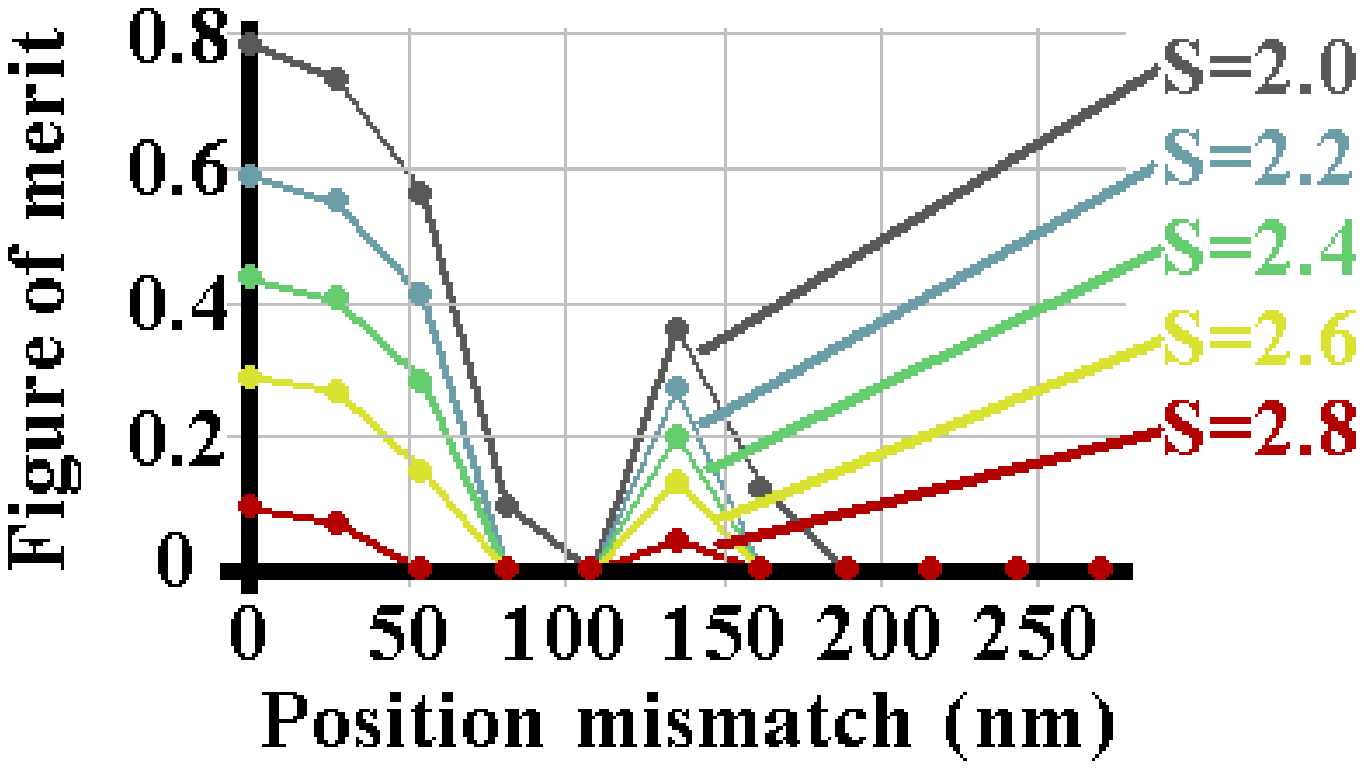} \\
   (a) & (b) & (c)
   \end{tabular}
   \caption{(a): Normalized $\beta$ factor as a function of the position mismatch of the dot. The red (blue) curve corresponds to the X (Y) polarization. (b): Figure of merit ($r=T_1^{bulk} \delta\omega / (\hbar F_p^{max})$) as a function of the position mismatch for various values of the Bell test: S=2, S=2.2, S=2.4, S=2.6 and S=2.8. }
   \label{fig:Excentre}
   \end{center}
\end{figure}

The position mismatch implies a breaking of the $C_6$ symmetry. The position of the dot will be identified by a direction $X$, as shown on figure \ref{fig:Excentre}(a), the $Y$ direction being orthogonal to the $X$ direction. The two polarization modes of the cavity remain unchanged. Therefore the sustained modes of the cavity will be described in the basis ($X$,$Y$) and no more in the ($H$,$V$) basis.
Due to the mismatch with the cavity modes, the dipole will
preferentially excite one of the modes ($X$ or $Y$ polarized) leading
to an unbalance of the fraction $\beta_i$ ($i\in[X,Y]$) of
spontaneous emission in the cavity mode. Inevitably, this will in
turn impact the entanglement visibility.

Figure \ref{fig:Excentre}(b) depicts the normalized $\beta_i$ factor
as a function of the position mismatch along the X direction for
both polarizations. The fraction of spontaneous emission for both
polarizations ($\beta_X$ and $\beta_Y$) at null mismatch; this permits us to normalize the $\beta_i$ to $1$ when the dot is
centered, and use in our simulations the extrapolated amplitude of the corresponding
mode divided by the amplitude of the whole field at the maximum of the excitation, in order to deduce the evolution of its $\beta_i$.

The effect of an asymmetric Purcell factor on the Bell's inequality
is modeled in the annexe of this paper. If we want to reach some
value $S$ of the Bell test whereas the dot is misplaced, this puts a
maximal limit to a figure of merit defined as the adimensional
ratio $r=T_1^{bulk} \delta\omega / (\hbar F_p^{max})$ where
$T_1^{bulk}$ is the bulk lifetime of the dot, 2$\hbar \delta\omega$ the excitonic energy 
splitting and $F_p^{max}$ the maximal Purcell effect of the cavity
(at zero mismatch). Entanglement visibility increases when $r$ tends to zero. This figure of merit only depends on the shortest lifetime that can be obtained in the dot-cavity system and on the quantum beat period between the two exciton states. Figure \ref{fig:Excentre} (c) depicts the
evolution of this figure of merit for different $S$ values,
calculated using the model introduced in the Annexe. For example, in the case of a centered dot with an excitonic bulk lifetime of $1$ ns, an excitonic energy splitting of about $2\mu eV$, submitted to a maximal Purcell factor of 10 and not subjected to incoherent processes,  the figure of merit $r=0.3$ allows $S$ to reach a value above $2.6$. Conversely, the Bell's inequality is hardly violated ($S=2$) if the same dot is about 70 nm away from the center of the cavity. For a more usual value of the excitonic splitting ($5 \mu eV$), the maximal mismatch enabling for Bell's inequality violation drops to 10 nm. 

\section{Conclusion}

In this paper we derived the Bell inequalities for a quantum dot
in a photonic crystal cavity with non-overlapping polarization modes. By analyzing the emission pattern of modified
H1 cavities, we demonstrate that it is possible to obtain both
high collection efficiencies (of the order of 22$\%$) and maximally overlapping modes while
keeping high Purcell factors. Finally we estimate that the
position of the quantum dot with respect to the cavity center has
to be more accurate than 50 nm, in order to implement an efficient quantum dot source of polarization entangled photons from the biexciton cascade.

\ack The authors would like to acknowledge L. Bernardi for
technical support. This work has been founded by the NanoEPR
project.

\section*{Annexe: effect of an assymetry in the Purcell effect}

The asymmetric branching ratio induced by a polarisation dependant
Purcell factor can be modelled as follows. Let the state of the
system (dot and optical fields) be
\begin{eqnarray}
\ket{\Psi(t)} &=& p_2(t) \ket{2;\emptyset;\emptyset}
+\sum_{u=H,V} \int d\omega_2 p_u(\omega_2,t) \ket{1_u;\vect{u},\omega_2;\emptyset} \nonumber\\
&& +\sum_{u=H,V} \int d\omega_2 d\omega_1
p_{uu}(\omega_1,\omega_2,t)
\ket{0;\vect{u},\omega_2;\vect{u},\omega_1}
\end{eqnarray}

\noindent where the first of the three entries within the ket
refers to the quantum dot's level, the two other entries refer to the
first and second emitted photons of polarisation $\vect{u}$ and
pulsation $\omega_i$ (i=1, 2 respectively) (see figure
\ref{fig:cascade}. We distinguish here the emission rates
$\gamma_1$ and $\gamma_2$ with respect to both polarizations $H$ and
$V$. We assume that incoherent processes are negligible, so that crossed terms combining horizontal and
vertical orientations disappear.)

The expressions of the $p_2$, $p_u$ and $p_{uu}$ coefficients are determined using the Wigner-Weisskopf approximation. Considering the system at long times ($t>>1/\gamma_1$ and $\gamma_2$) the terms $p_2(t)$ and $p_u(\omega_2,t)$ tend to zero, which gives a state that can be factorized into a radiative part $\ket{\Psi_R}$ and the fundamental source state $\ket{0}$. Considering no spectral filtering, the density matrix of the photon pair in the polarization basis is:
\begin{eqnarray}
\rho &=& \int d\omega_1 d\omega_2 \bra{\omega_2,\omega_1}\op{\Psi_R}{\Psi_R}\ket{\omega_2,\omega_1} \\
  &=& \sum_{u,v=H,V} \op{\vect{u}\vect{u}}{\vect{v}\vect{v}} \int d\omega_2 d\omega_1 p_{uu}(\omega_1,\omega_2,\infty) p_{vv}(\omega_1,\omega_2,\infty)^* \\
  &=& \frac{1}{2(1+2\delta F^2)} \left( \begin{array}{cccc}
    (\delta F+1)^3 & 0 & 0 & \frac{(1-\delta F^2)^2}{1-i g} \\
    0 & 0 & 0 & 0 \\
    0 & 0 & 0 & 0 \\
    \frac{(1-\delta F^2)^2}{1+i g} & 0 & 0 & (\delta F-1)^3
  \end{array} \right)
\end{eqnarray}
where we defined the relative difference of Purcell factors
$\delta F=(F_H-F_V)/(F_H+F_V)$ and the normalized splitting
$g=2\delta\omega/(\gamma_1^{bulk}(F_H+F_V))$. $\delta F$ is approximated from the ration between the modal coupling factors as: $(\beta_H-\beta_V)/(\beta_H+\beta_V)$. In the same way as
in \cite{Larque2008}, we deduce the expression $S$ of the Bell test,
from which we deduce data presented on figure \ref{fig:Excentre}(c).


\end{document}